\documentstyle{article}[12pt]

\def\a{\alpha}
\def\b{\beta}
\def\d{\delta}
\def\D{\Delta}

\def\g{\gamma}

\def\e{\epsilon}

\def\k{\kappa}
\def\l{\lambda}

\def\m{\mu}
\def\n{\nu}
\def\s{\sigma}

\def\r{\rho}
\def\t{\tau}

\def\car{{\cal R}}

\def\cv{{\cal V}}

\renewcommand{\@}[1]{\sqrt{#1}}

\renewcommand{\le}[1]{\label{#1}\end{eqnarray}}
\newcommand{\be}{\begin{equation}}
\newcommand{\ee}{\end{equation}}
\newcommand{\bea}{\begin{eqnarray}}
\newcommand{\eea}{\end{eqnarray}}
\newcommand{\nn}{\nonumber}

\def\nn{\nonumber\\}

\def\ffract#1#2{\raise .35 em\hbox{$\scriptstyle#1$}\kern-.25em/
\kern-.2em\lower .22 em \hbox{$\scriptstyle#2$}}

\def\half{{1\over2}\,}
\def\nonu{\nonumber \\{}}

\newcommand{\del}{\partial}

\begin{document}
\rm\large \null\vskip-24pt
\begin{flushright}
UCI-TR-2005-19\\
{\tt hep-th/0505155}
\end{flushright}
\vskip0.1truecm
\begin{center}
\vskip .5truecm {\Large\bf
On a supersymmetric completion of the $R^4$ term in \\
type IIB supergravity } \vskip 1truecm {\large\bf Arvind Rajaraman
  \footnote{ e-mail: {\tt arajaram@uci.edu}}
}\\
\vskip .5truecm {\it Department of Physics and Astronomy,
University of California, \\
Irvine, CA 92697, USA}
\end{center}
\vskip .3truecm
\begin{center}
{\bf \large Abstract}
\end{center}
We examine the question of the supersymmetric completion of the $R^4$
term in type IIB supergravity by using superfield methods. We show
that while there is an obstruction to constructing the full action, a
subset of the terms in the action can be consistently analyzed
independent of the other terms, and these can be obtained from the
superfield. We find the complete type IIB action involving the
curvature and five-form field strengths.

\setcounter{equation}{0}

\vskip 2 cm\section{Introduction}

At low energies, string theory can be reduced to an effective field
theory of the massless modes. All string theories have a massless
graviton, and to leading order, the action for this field is the
Einstein-Hilbert action. Supersymmetric type II strings have a large
number of fields in addition to the graviton, and the complete
two-derivative action for these fields is the $N=2$ supergravity
action in ten dimensions.

The effective action also contains an infinite series of higher
derivative terms, suppressed by powers of the string scale $\alpha'$,
and the complete action has the form \bea \k^2S=S_2+(\alpha')^4S_8
+(\alpha')^5 S_{10}+\dots \eea where $S_n$ contains terms with $n$
derivatives. $S_2$ is the supergravity action. The leading correction
in type II theories is the eight-derivative action, which contains
the famous $R^4$ term \cite{GW,Gross:1986mw}\bea S_{8;R^4}=\int
d^{10}x\ t^8t^8R^4 \eea This term occurs in all string theories and
in the eleven-dimensional theory.

There are also several other terms at the eight derivative level
which involve the other fields of the theory (the Kalb-Ramond field,
the dilaton and the Ramond-Ramond fields). These terms are believed
to be related to the $R^4$ term by supersymmetry. Here we will
discuss a procedure for finding (some of) these terms. \vskip 1 cm

There are  many reasons that one wishes to know the full action at
the eight-derivative level.

At the basic level, knowledge of these terms will tell us a lot more
about actions with maximal supersymmetry, which may lead to
fundamental understandings like the off-shell nature of the theory.

From a phenomenological viewpoint, there has been a lot of interest
in flux compactifications, where fluxes are turned on in the internal
Calabi-Yau manifold (e.g. \cite{Kachru:2003aw}). The potential for
moduli in this background can be efficiently computed in the low
energy effective theory, and can be used to gain information about
stable compactifications at large radius. However, one needs to know
the full action including the Ramond-Ramond (RR) field strengths.

Another place where the full effective action is required is for
computing $\alpha'$ corrections in Anti-de-Sitter (AdS) backgrounds,
for applications to the AdS/CFT correspondence. These can be applied
to find corrections to black hole entropy, or to correlation
functions.

\vskip 1 cm

Despite these motivations, it has not been possible so far to
determine the complete eight derivative action. Several different
approaches have been tried:
\begin{itemize}
\item The action can be computed by evaluating all the relevant
string diagrams, and extracting the low energy action (see for
example \cite{GW,Gross:1986mw,Peeters:2001ub,ampl1}). The $R^4$ term
can be found in this way. A related approach is to use sigma model
techniques \cite{GVZ}. Another related approach was tried in
\cite{moreprevious}.

\item One can attempt to use dualities to generate terms involving RR
fields, using the known terms involving NSNS fields.

\item  It is believed that the eight-derivative action is completely
determined by supersymmetry alone. One can therefore attempt to
construct the action by using the Noether method to generate terms
step by step until supersymmetry is satisfied. This has been
attempted for the heterotic string action
\cite{Romans:1985xd,Bergshoeff:1989de,deRoo:1992zp,Sch}.

\item If the complete superfield can be found, then the action can be
written as an integral over one-half of superspace. This has also
been attempted for the heterotic string in \cite{Nilsson:1986rh}, and
discussed for the maximally supersymmetric theories
\cite{Peeters:2001qj} (see also \cite{Deser:nt}).
\end{itemize}

Each of these approaches has serious difficulties.

\begin{itemize}
\item String diagrams contain much more information than just the
eight-derivative terms. One needs an effective way of extracting the
low energy limit without doing the entire computation. Furthermore,
once we get to five-point amplitudes and beyond, we have to worry
about extracting contributions to the amplitude involving the
exchange of massless fields, for example those coming from a
combination of the four-point eight derivative amplitude and a tree
level three-point interaction. Furthermore, the plethora of fields in
the supergravities means that many amplitudes need to be computed.
Sigma model techniques also require intense computational effort.

\item Dualities do not determine the action. In general, dualities
exchange an action at one value of a modulus with another action at
another value of the modulus. While this constrains the way moduli
can appear in any particular term in the action, it does not, in
general, relate different terms in the action.

\item The Noether method is completely general, but the vast number
of fields and the plethora of possible terms make it impractical to
use this method directly in ten dimensional supergravity.

\item The superfield approach is by far the most promising.
Unfortunately, there is a no-go result \cite{deHaro:2002vk}: it is
known that the eight-derivative action of type IIB supergravity
cannot be written as the integral over one-half of superspace of a
scalar superfield.
\end{itemize}

In this paper, we shall show that the superfield approach can be
modified to obtain some information about the effective action.

\vskip 1 cm

Previous work on the eight derivative terms has produced many
important results \cite{Green:1999qt,Kiritsis:1997em}. Most
importantly, several nonrenormalization theorems are known which
strongly restrict the moduli dependence of the eight-derivative
terms. We shall review some of these results further on.

Secondly, a lot of work has been done on the superfield approach to
type IIB supergravity, starting with the seminal work of \cite{HW}.
Particularly important for us will be the work of
\cite{deHaro:2002vk}. These authors applied the superfield approach
to constructing the eight derivative terms in type IIB supergravity,
and showed that there was an obstruction to constructing a viable
supersymmetric action. We shall review their approach, and show how
the obstruction they found can be (partially) avoided.

\vskip 1 cm

Our first new result will be the construction of the quartic terms in
type IIB supergravity. This construction can be done using the
linearized superfield, and has not appeared in the literature to our
knowledge (although we believe that this construction is well known
to the experts.) We shall also describe the moduli dependence of
these terms.

We then go beyond the quartic level. We do this by the superfield
construction. The supersymmetric measure does not exist in general,
but we shall show that if we restrict ourselves to a well-defined
subset of the terms, the measure can be constructed, and
supersymmetric actions can be constructed. Our main result will be
the complete type IIB action involving curvature and five-form field
strengths alone.

We should emphasize that the specific form for this action has been
found before in the literature \cite{previous}. These papers also
used the integral of the superfield to find the action.
Unfortunately, the analysis of \cite{deHaro:2002vk} has shown that
the full action cannot be written using superfields, and this {\it a
priori} invalidates the results in these previous papers. The present
paper shows that, in fact, the results obtained by these previous
authors was correct, despite the no-go theorem of
\cite{deHaro:2002vk}. This paper can hence be taken partly as a
justification of the previous results in \cite{previous}.

We will finally close with a discussion of our results and make a
conjecture as to the extension of our results to find the other terms
in the action.

\section{Type IIB in components}

We now discuss type IIB supergravity. This was first discussed in
\cite{S}; we will follow the conventions of \cite{deHaro:2002vk}.

The field content of Type IIB supergravity consists of the vielbein
$e_\mu^{~a}$, a complex two-form field $a_{\mu\nu}$, a real four form
field $a_{\mu\nu\rho\s}$, and a complex scalar $a$. The fermionic
fields are a graviton $\psi_\mu$ and a dilatino $\l$.

The scalar $a$ transforms in a nonlinear representation under the
U-duality group $SU(1,1)$. To represent the symmetry linearly, we
introduce the new fields \bea {\cal V}= \left(
\begin{array}{cc} u&v\\v^*&u^*\end{array}\right)\eea  with
$uu^*-vv^*=1$. The scalars $u,v$  parametrize the $SU(1,1)/U(1)$
coset manifold.  $SU(1,1)$ acts on the left, and $U(1)$ acts from the
right \be \label{su11} \cv'= \left(
\begin{array}{cc}
z & w \\
w^* & z^*
\end{array}
\right) \left(
\begin{array}{cc}
u & v \\
v^* & u^*
\end{array}
\right) \left(
\begin{array}{cc}
e^{-i \Sigma} & 0 \\
0 & e^{i \Sigma}
\end{array}
\right)~. \ee

The physical field $a$ is given by $a={v\over u^*}$. It is invariant
under $U(1)$, and transforms under $SU(1,1)$ as \be a' = {z a + w
\over w^* a + z^*} \ee

Furthermore, the axion $C_0$ and dilaton $\phi$ of type IIB string
theory can be defined by \be \tau=C_0 + i \exp(-\phi)=i{1-a \over
1+a}\ee

To represent the scalar kinetic terms, we introduce the $SU(1,1)$
invariant objects \bea p=u^* d v - v d u^*, \qquad q={1 \over 2 i}
(u^* d u - v d v^*) \eea $q$ transforms as a $U(1)$ connection.

The gauge field strengths are defined with weight $n$ i.e. \bea {\cal
F}_{abc}=3\del_{[a}a_{bc]}~~~~~~~~~~~~~~~~~~~~~~~~~
\\
g_{abcde}=5\del_{[a}a_{bcde]}-10i(a^*_{[ab}{\cal
F}_{cde]}-a_{[ab}{\cal F}^*_{cde]})
\eea as is the gravitino field strength \bea
\psi_{ab}=2\del_{[a}\psi_{b]}
\eea
 The five-form field strength $g_5$ is self-dual, and also invariant under
 $SU(1,1)$. The field strengths ${\cal F}_3$ transform as a doublet
 under  $SU(1,1)$;
  we therefore define the $SU(1,1)$ invariant
3-form field strengths \bea (f_3^*,f_3)=({\cal F}^*_3,{\cal
F}_3){\cal V} \eea Under the local U(1), the complex field strength
$f_3$ then has charge 1.

In terms of these fields the supersymmetry transformation laws are

\be \d e_{\m}{}^a = - i \left( (\zeta^* \g^a \psi_\m ) + (\zeta \g^a
\psi^*_\m ) \right) \ee \bea \d \psi_\m &=& D_\m \zeta - \frac{3}{16}
\hat{f}_{\m ab} \g^{ab} \zeta^* + \frac{1}{48} \hat{f}^{bcd} \g_{\m
bcd} \zeta^* - \frac{1}{192} i \hat{g}_{\m abcd} \g^{abcd} \zeta \nn
&&+ \frac{1}{16} i \left[ - \frac{21}{2} (\l^* \g_\m \l) +
\frac{3}{2} (\l^* \g^a \l) \g_{\m a} + \frac{5}{4} ( \l^* \g_{\m ab}
\l) \g^{ab} - \frac{1}{4} (\l^* \g^{abc} \l) \g_{\m abc} \right. \nn
&&\left. - \frac{1}{16} (\l^* \g_{\m abcd} \l) \g^{abcd} \right]
\zeta - (\zeta^* \g^a \psi_\m^*) (\g_a \l) + (\psi^*_\m \l) \zeta^* -
(\zeta^* \l) \psi^*_\m \eea \be \d u = 2 (\zeta^* \l^*) v \qquad \d v
= - 2 (\zeta \l) u \ee \be \d \l = \frac{1}{24} i \hat{f}_{abc}
\g^{abc} \zeta + \frac{1}{2} i \hat{p}_a \g^a \zeta^* \ee \be \d
(a^*_{\m\n}, a_{\m\n}) = - \left( (\zeta \g_{\m\n} \l^*)  + 2 i
(\zeta^* \g_{[\m} \psi^*_{\n]} ) , - (\zeta^* \g_{\m\n} \l) + 2 i
(\zeta \g_{[\m} \psi_{\n]}) \right) \cv^{-1} \ee \be \d a_{\m\n\r\s}
= - 4 (\zeta \g_{[\m\n\r} \psi^*_{\s]} ) + 4 (\zeta^* \g_{[\m\n\r}
\psi_{\s]} ) + 12 i \left( a_{[\m\n} \d a^*_{\r\s]} - a^*_{[\m\n} \d
a_{\r\s]} \right) \ee

Here \bea D_\m\e= \del_\m\e-{1\over 4}\omega_{\m}^{~bc}\g_{bc}\e\eea
and the spin connection is defined as \bea
2\omega_{\m}^{~bc}=-e^{b\s}(\del_\m e_{\s}^c-\del_\s e_{\m}^c)
+e^{c\s}(\del_\m e^b_{\s}-\del_\s e_{\m}^b)+e^{b\s} e^{c\rho}
e_\mu^e(\del_\s e_{e\rho}-\del_\rho e_{e\s}) \eea

We have also defined the supercovariant expressions \bea && \hat{p}_a
= p_a + 2 (\psi_a \l) \nn && \hat{f}_{abc} = f_{abc} - 3 (\psi^*_{[a}
\g_{bc]} \l) - 3 i (\psi_{[a} \g_b \psi_{c]}) \nn && \hat{f}_{abcde}
= f_{abcde} + 20 (\psi^*_{[a} \g_{bcd} \psi_{e]}) \eea

The U(1) charges of every field are now fixed by the consistency of
the supersymmetry transformations. We find that under the local U(1),
the charges of the fields are \bea e_\mu^{a},g_5: 0\qquad \psi_\m,
\e: {1\over 2}\qquad  f_3:1 \qquad \l: {3\over 2}\qquad p_a: 2\eea

\section{The $R^4$ terms}

The eight-derivative effective action $S_8$ is a sum of several
terms, each of which has a moduli-dependent coefficient. For type
IIB, the effective action has the generic form \bea
S_8=f_{R^4}(\tau,\tau^*)R^4+f_{R^2f_3^2}(\tau,\tau^*)R^2f_3^2+
f_{R^2\l\l}(\tau,\tau^*)R^2\l\l+\dots \eea where we have shown a few
of the many possible terms.

The structure of the $R^4$ terms can be made more explicit.
Calculations in type IIB string theory \cite{GSW} show that the
complete action involving the curvature alone is \bea S_{8;R^4}=
f_{R^4}(\tau,\tau^*)(t^{abcdefgh}t_{ijklmnop}+\e^{IJabcdefgh}\e_{IJijklmnop}
)R_{ab}^{~~ij}R_{cd}^{~~kl}R_{ef}^{~~mn}R_{gh}^{~~op} \eea where
$t^{abcdefgh}$ is defined in Appendix 9.A of \cite{GSW}. The
structure of the remaining terms have not been explored in as much
detail.

The requirement of $U(1) $ invariance strongly constrains the
moduli-dependent coefficients. For example, $R^4$ is invariant under
the local $U(1)$ symmetry, and so $f_{R^4}(\tau,\tau^*)$ must also be
invariant under the  symmetry. Similarly, ${R^2f_3^2}$ has a charge
2, and so $f_{R^2f_3^2}(\tau,\tau^*)$ should have charge $-2$ under
the symmetry.

The effective action can therefore be separated into different pieces
according to the U(1) transformation law of the moduli-independent
terms \bea S_8=S_{8;0}+S_{8;1}+\dots +S_{8;24} \eea where the
$S_{8;i}$ have the schematic form \bea \label{actionform}
S_{8;0}=f^{(0,0)}(\tau,\tau^*)(R^4+R^2\psi^*\psi+(g_5)^8+R^2f_3f_3^*+\dots)
\\
 S_{8;1}=f^{({1\over 2},-{1\over 2})}(\tau,\tau^*)(R^2\psi\psi+\dots)
 \\
 S_{8;2}=f^{(1,-1)}(\tau,\tau^*)(R^2\psi\l+R^2f_3f_3+\dots)
 \\
 \dots
 \\
 S_{8;24}=f^{(12,-12)}(\tau,\tau^*)(\l^{16})
 \eea
The coefficients have been written in terms of the modular forms
$f^{(w,-w)}(\tau,\tau^*)$. Also, we have made an assumption here that
the $U(1)$ transformation uniquely determines each moduli-dependent
coefficient up to a constant. This assumption was justified by the
detailed analysis of \cite{Green:1999qt}.

 Under a $SL(2,Z)$
transformation \bea \tau\rightarrow {a\tau+b\over c\tau+d}\eea these
functions transform as \bea f^{(w,-w)}(\tau,\tau^*)\rightarrow
f^{(w,-w)}(\tau,\tau^*)(c\tau+d)^w(c\tau^*+d)^{-w}\eea

The explicit form of the modular forms can be determined by
supersymmetry \cite{Green:1999qt}. We define the action of the
modular covariant derivative on non-holomorphic forms
$f^{(w,-w)}(\tau,\tau^*)$ as \bea D_w=i\left(\tau_2{\del\over \del
\tau}-i{w\over 2}\right)\eea Then supersymmetry implies that \bea
\label{fwformula}f^{(w+1,-w-1)}(\tau,\tau^*)=D_w
f^{(w,-w)}(\tau,\tau^*) \eea

Furthermore, we find that $f^{(w,-w)}(\tau,\tau^*)$ are
eigenfunctions of the Laplacian on the $SL(2,Z)$ space. In
particular, one finds \bea \nabla^2 f^{(0,0)}(\tau,\tau^*) = {3\over
4}f^{(0,0)}(\tau,\tau^*)\eea

For weak coupling, $f ^{(0,0)}(\tau,\tau^*)$ should have a power law
behavior. This fixes $f ^{(0,0)}(\tau,\tau^*)$ to be the Eisenstein
series of order ${3\over 2}$ \bea f^{(0,0)}(\tau,\tau^*)
={\sum_{m,n\neq (0,0)}} {\tau_2^{3\over2}\over |m+n\tau|^{3}} \eea
The other $f^{(w,-w)}(\tau,\tau^*)$ are then determined by
(\ref{fwformula}).

\section{Type IIB in superfields}

We now turn to the superfield formulation of type IIB supergravity.
This section follows the formulation of \cite{deHaro:2002vk} very
closely.

The superspace formulation of  type IIB supergravity was first
constructed in \cite{HW}. One introduces a Grassmann variable
$\theta^\a$ which is a 16-component Weyl spinor of $SO(9,1)$. The
superspace coordinates are then
$(x^\mu,\theta^\a,(\theta^*)^{\bar{\a}})$. The covariant derivatives
satisfy the algebra \bea [D_A,D_B\}=-T_{AB}^{~~C}D_C+{1\over
2}R_{ABC}^{~~~~D}L_D^{~C} +2iM_{AB}\k \eea where $T_{AB}^{~~C}$ is
the torsion, $R_{ABC}^{~~~~D}$ is the curvature, and $M_{AB}$ is the
U(1) curvature (the explicit values of these tensors can be found in
\cite{deHaro:2002vk}). The super-Jacobi identities then produce a
large set of relations.

To obtain the field content of type IIB supergravity, we need to
impose constraints on the superspace. After the imposition of the
constraints, the relations obtained from the super-Jacobi identities
become nontrivial and need to be solved. This was done in \cite{HW},
where it was shown that the fields of type IIB supergravity could be
obtained from a chiral superfield $V$ satisfying   \bea
D^*_{\alpha}V=0\eea

 The super-Jacobi identities  provide several further constraints on the superfield.
These constraints
 were summarized in \cite{HW,deHaro:2002vk}.

All components of the superfield $V$ can be obtained by solving these
constraints. The first few components are found to be
\cite{HW,deHaro:2002vk} \bea
V| &=& v \\
D_\a V| &=& - 2 u \l_\a \\
D_{[\a} D_{\b]} V| &=& {i \over 12} u \g_{\a \b}^{abc}
\hat{f}_{abc} \\
D_{[\g} D_\b D_{\a]} V| &=& \frac{i}{12}  u \g^{abc}_{\b \a} \left\{
-\frac{1}{32} \left( \g_{abcdef} \hat{f}^{* def} + 3 \hat{f}^*_{
[a}{}^{de}\g_{bc]de} + 52 \hat{f}^*_{ [ab}{}^{d} \g_{c]d} + 28
\hat{f}^*_{ abc} \right)_\g{}^{\e} \l_\e \right. \nonu &+& \left. 3
\hat{p}_{[a} (\g_{bc]})_\g{}^\e \l^*_e + 3 i (\g_a)_{\g \e}
\hat{\psi}_{bc}^\e \right\}. \eea

The bosonic part of the fourth component is given by \be D_{[\d}
D_{\g} D_{\b} D_{\a]} V| = u \g^{abc}_{[\b\a}\g^{def}_{\g \d]}
\car_{abcdef} \ee where \be \label{car} \car_{abcdef}
=\frac{1}{16}(g_{ad} c_{bcef} - \frac{i}{6} D_b g_{acdef})
-\frac{1}{1536} (3 g_{bafmn} g_{ced}{}^{mn} - g_{abcmn}
g_{def}{}^{mn})
 + f^*_3 f_3 {\rm\ terms}~.
\ee where $c_{bcef}$ is the Weyl tensor.

Finally, the supersymmetry variations of the component fields can be
obtained by using the equation for the variation of the superfield
under supersymmetry transformations \bea \d_\xi V={\xi}^*D_\a^* V
+{\xi}D_\a V \eea

The superfield $U^*$ is also chiral; it is however not independent
since it satisfies \bea UU^*-VV^*=1\eea Any function of $V, U^*$ is
also chiral.

\section{From superfields to the action} Now that we have constructed superfields,
one can attempt to use them to construct supersymmetric actions. To
get an eight derivative action, we integrate over one-half of
superspace.

\subsection{The linear case}
We will first try this with the linearized superfield.  We first
gauge fix the $U(1)$ symmetry by assuming that the vacuum solution is
given by $u=1, v=0$, i.e. \bea {\cal V}_0= \left(
\begin{array}{cc} 1&0\\0&1\end{array}\right)\eea Linearized
perturbations about this background value will be represented by \bea
{\cal V}_{lin}= \left(
\begin{array}{cc} 1&v\\v^*&1\end{array}\right)\eea
At the linearized level, we can set $v_{lin}=a_{lin}$.

The first few components of the linearized superfield can now be
immediately obtained from the above formulae \bea
V_{lin}| &=& v \\
D_\a V_{lin}| &=& - 2  \l_\a \\
D_{[\a} D_{\b]} V_{lin}| &=& {i \over 12}  \g_{\a \b}^{abc} {f}_{abc}
 \\
D_{[\g} D_\b D_{\a]} V_{lin}| &=& -\frac{1}{4}   \g^{abc}_{\b \a}
(\g_a)_{\g \e} {\psi}_{bc}^\e
\eea while the bosonic part of the fourth component is\be D_{[\d}
D_{\g} D_{\b} D_{\a]} V_{lin}| = \frac{1}{16}
\g^{abc}_{[\b\a}\g^{def}_{\g \d]}
 (g_{ad} R_{bcef} - \frac{i}{6} D_b
g_{acdef}) \ee

To get the $R^4$ action, we integrate $V^4$ over one half of
superspace to get \cite{deHaro:2002vk}\bea  \int d^{16}\theta
(V_{lin})^4= (t^{abcdefgh}t_{ijklmnop}+\e^{IJabcdefgh}\e_{IJijklmnop}
)R_{ab}^{~~ij}R_{cd}^{~~kl}R_{ef}^{~~mn}R_{gh}^{~~op} +\dots\eea This
is in agreement with the string calculation, which is an indication
that we may be on the right track.

To fill out the rest of the factors in (\ref{actionform}), we
construct the action \bea \label{linaction} S_{8;lin}=\int d^{10}x
\sqrt{g} f^{(0,0)}(\tau,\tau^*)\int d^{16}\theta (V_{lin})^4\eea This
expression is not completely supersymmetric; however all variations
containing at most four fields cancel.

 The other terms from the expansion of the action
will then produce the quartic action of type IIB supergravity.

\subsection{The nonlinear case and a failure}

We now review the results obtained by \cite{deHaro:2002vk} for the
nonlinear action.

When we try to go beyond the quartic action, we will need the full
nonlinear superfield. In addition we need a supersymmetric measure;
the supersymmetric analogue to the $\sqrt{g}$ factor. The suggested
form of the eight-derivative action is then \bea \label{act1}
S_8=\int d^{10}x \int d^{16}\theta \Delta W[V,U^*] \eea where
$\Delta$ is by definition a superfield whose lowest component is \bea
\label{Delta0}\Delta|_{\theta=0}=\sqrt{g} \eea $\Delta$ is to be
constructed order by order by requiring that the action be
supersymmetric. (When we were considering the linear action, we did
not need to worry about constructing $\D$, since it only contributes
to higher order terms.)

The action can be written\bea  S_8 &=& \int d^{10} x\,
\e^{\a_1..\a_{16}} \sum_{n=0}^{16} {1 \over n! (16 -n)!} D_{\a_1}
..D_{\a_{n}} \D|\, D_{\a_{n+1}}...D_{\a_{16}} W| \nn &=& \int d^{10}
x \sum_{n=0}^{16} {1 \over n!} D_{\a_1} ..D_{\a_{n}} \D|\,
D^{16-n,\a_1 ..\a_n} W| \eea where we have introduced the notation
\be \label{def} D^{16-n,\a_1 ..\a_n} W = {1 \over (16-n)!}\,
\e^{\a_1..\a_{16}} D_{\a_{n+1}}...D_{\a_{16}} W~. \ee

Invariance of the action under supersymmetry requires \be
\label{daction} \d S = \int d^{10} x \sum_{n=0}^{16} \frac{1}{n!}
\left( \d D_{\a_1} ..D_{\a_{n}} \D| D^{16-n,\a_{1}..\a_{n}} W| +
D_{\a_1} ..D_{\a_{n}} \D| \d D^{16-n,\a_{1}..\a_{n}} W| \right)=0.
\ee

The first projection of $\D$ is determined by equation
(\ref{Delta0}). The next projection can be determined by requiring
the cancellation of variations containing $D^{16}W$ and $\zeta$. This
yields \cite{deHaro:2002vk} \bea
D_\a\Delta|_{\theta=0}=-ie\g^c_{\a\b}\psi^{*\b} \eea Similarly, the
cancellation of terms proportional to $D^{15}W$ determines \bea
[D_\a,D_\b]\Delta|_{\theta=0}= \frac{1}{12} i e \g^{abc}{}_{\a \b}
f^*_{abc} + O[\psi^* \psi^*,\l^* \psi]  \eea

At the same time, the terms proportional to $\zeta^*$ need to cancel
as well, and this has to happen automatically for this construction
to work. As it turns out, the terms containing $D^{16}W$ and
$\zeta^*$ do cancel, but at next order the cancellation does not work
\cite{Howe,deHaro:2002vk}. The uncancelled term is \be
\label{obstruction}\d S = \half \int d^{10} x e \zeta^{*\a}
T_{\bar{\a} \bar{\d}}^\g T_{\g \b}^{\bar{\d}}  D^{\b, 15} W| \ee

This then implies that the eight-derivative action cannot be written
as an integral over one-half of superspace.

\section{A second attempt at an action}

This analysis suggests that we should give up the idea of reproducing
the complete action from this integral, which as we have seen is a
hopeless task. Instead we can try to use the superfield to reproduce
a {\it subset} of the terms in the action. In fact, we will now show
that the superfield construction can be used to find the complete
effective action containing just the curvature and five-form field
strength $g_5$.

\vskip 1 cm

First, we must show that there is a consistent way to restrict
ourselves to this subset of terms. The $U(1)$ symmetry will help us
here.

The curvature and five-form field strength $g_5$ are both uncharged
under the $U(1)$ symmetry. Hence all the terms containing only these
two objects are in $S_{8;0}$, and will occur multiplied by
$f^{(0,0)}(\tau,\tau^*)$. The bosonic terms containing only $R, g_5$
then have the schematic form \bea S_{8;0; R,g_5}= \int d^{10} x
f^{(0,0)}(\tau,\tau^*)(R^4+g_5^8+\dots )\eea

Under a supersymmetry variation, these terms produce a huge set of
variations that need to be cancelled. We will  consider the subset of
variations which have at most one fermion field, and where we set
$\del{\tau^*}=\del{\tau}=\l=a_2=0$. We can then ignore the variation
of $f^{(0,0)}(\tau,\tau^*)$.

The remaining variations are of the schematic form \bea \d S_{8;0;
R,g_5}=\int d^{10} x
f^{(0,0)}(\tau,\tau^*)(R^3\psi^*\e+(g_5)^7\psi^*\e+\dots )\eea These
can be cancelled by variations coming from terms containing fermion
bilinears which are of the schematic form \bea S_{8;0;
R,g_5,\psi^*,\psi}=\int d^{10} x
f^{(0,0)}(\tau,\tau^*)(R^2\psi^*\psi+(g_5)^7\psi^*\psi+\dots )\eea We
note that it is essential for the gravitino terms to have a $\psi^*$
as well as $\psi$ in order to cancel the $U(1)$ charge.

\vskip 1 cm

 We now argue that we can consistently restrict
ourselves to this subset of terms in the action viz. $S_{8;0; R,g_5}$
and $S_{8;0; R,g_5,\psi^*,\psi}$, as long as we only look for the
cancellation of the subset of the variations $\d S_{8;0; R,g_5}$
discussed above.


The reason this is not obvious is that the variations of a term
linear in a field (say $a_2$) includes terms independent of $a_2$. So
one might expect that in general, a cancellation of the supersymmetry
variations $\d S_{8;0; R,g_5}$ which satisfy
$\del{\tau^*}=\del{\tau}=\l=a_2=0$ will require us to consider terms
in the action which are linear in these fields.

In this case, the $U(1)$ symmetry helps us out.  $U(1)$ invariance of
the bosonic terms implies that the terms multiplied by
$f^{(0,0)}(\tau,\tau^*)$ cannot be linear in the fields
$\del{\tau^*},\del{\tau},\l,a_2$, which are all charged under the
$U(1)$. It is possible to have terms linear in these fields if they
are multiplied by a compensating factor $f^{(w,-w)}(\tau,\tau^*)$
with the appropriate $U(1)$ charge, but these terms cannot contribute
to the cancellation of the variations in $\d S_{8;0; R,g_5}$.

We can have terms containing fermion bilinears which are linear in
$a_2$, where the $U(1)$ charge is cancelled by having $\psi\psi$
instead of $\psi^*\psi$. However the variation of $a_2$ will then
produce variations which are trilinear in fermion fields, and which
again do not contribute to the cancellation of $\d S_{8;0; R,g_5}$.

Accordingly, we can set $\del{\tau}=\l=a_2=0$ for all terms in
$S_{8;0; R,g_5}$ and $S_{8;0; R,g_5,\psi^*,\psi}$ contributing to
cancel $\d S_{8;0; R,g_5}$. In other words, we can restrict ourselves
to the subset of terms $S_{8;0; R,g_5}$ and $S_{8;0;
R,g_5,\psi^*,\psi}$.

\vskip 1 cm

We can now attempt to find the precise form of the action by
requiring that the supersymmetry variations coming from $S_{8;0;
R,g_5}$ and $S_{8;0; R,g_5,\psi^*,\psi}$ cancel. This would be an
extremely tedious procedure. Instead we shall see that the superfield
offers a quick way to reproduce these terms in the action.

As we have seen, we can set $\del{\tau}=\l=a_2=0$. The  $U(1)$
structure then ensures that the terms we are looking for all occur in
the superfield only in the third, fourth and fifth components i.e.
with a factor of $\theta^3, \theta^4$ or $\theta^5$. In particular
the bosonic terms are all found in the $\theta^4$ component.

To get a 8-derivative term, we should consider the action \bea
S_{8}=\int d^{10}x  g(\tau,\tau^*)\int d^{16}\theta \Delta V^4\eea
which is the natural extension of the linearized action
(\ref{linaction}).

To obtain bosonic terms, we should look at the $\theta^4$ component
in V. The 16 $\theta$ are then already saturated from the $V^4$ term.
For the terms bilinear in fermions, at least 15 $\theta$ must be
taken from the $V^4$ term.

Hence to construct the action, we only need the first two components
of $\Delta$, i.e. $\Delta|_{\theta=0}\equiv \sqrt{g}$ and
$D_\a\Delta|_{\theta=0}$. We do not need the other components of the
measure, as long as we are restricting ourselves to this particular
subclass of terms. That is, we may truncate the action to \bea
\label{action} S = \int d^{10} x\, {1 \over
16!}g(\tau,\tau^*)\e^{\a_1..\a_{16}} \left(\sqrt{g}\,
D_{\a_{1}}...D_{\a_{16}} W| +16 D_{\a_1}\D|\,
D_{\a_{2}}...D_{\a_{16}} W|\right) \eea where  we have set $W\equiv
V^4$. We can now consider the variations of this action. \vskip 1 cm

We are setting $\del{\tau}=\l=a_2=0$. Furthermore, we are considering
variations with at most one fermion field. In this case, we can set
$D^{n}W|=0$ in the supersymmetry variations for all $n\leq 14$.
 We then only need to cancel the
variations proportional to $D^{16}W$ and  $D^{15}W$.

Now, as we discussed above, the components of $\Delta$ have already
been computed in \cite{deHaro:2002vk},
yielding  \bea \label{Deltaeqn}\Delta|_{\theta=0}=\sqrt{g} \\
D_\a\Delta|_{\theta=0}=-ie\g^c_{\a\b}\psi^{*\b} \eea Furthermore, it
was shown that the variations proportional to $D^{16}W$ and $D^{15}W$
do cancel up to the obstruction shown in equation
(\ref{obstruction}). However, the torsion factor is \bea T_{\g
\b}^{\bar{\d}}&=& (\g^{a})_{\g \b} (\g_a)^{\d \t} \l^*_\t - 2
\d_{(\g}^{\d} \l^*_{\b)}    \eea which vanishes if we set
$\del{\tau}=\l=a_2=0$ in the variations. This means that the
obstruction vanishes, and the variations $\d S_{8;0; r,g_5}$ indeed
cancel in the action (\ref{action}).

\vskip 1 cm

Finally, we need to fix the moduli-dependent function
$g(\tau,\tau^*)$. To do this, we require that the action reproduce
the known $R^4$ term. Noting that the bosonic part of the fourth
component is given by \be D_{[\d} D_{\g} D_{\b} D_{\a]} V| =
\frac{1}{16}u \g^{abc}_{[\b\a}\g^{def}_{\g \d]}  (g_{ad}
R_{bcef}+\dots)\ee we find that up to an overall constant \bea
g(\tau,\tau^*)={1\over u^4}f^{(0,0)}(\tau,\tau^*) \eea This
reproduces the $R^4$ term. The supersymmetric completion including
$g_5$ terms is then given by (\ref{action}).

The bosonic part of the action can be immediately written down; up to
an overall constant\bea \label{bosact} S_{8;0; R,g_5}= \int d^{10} x
f^{(0,0)}(\tau,\tau^*)\times~~~~~~~~~~~~~~~~~~~~~~~~~~~~~~~~~~~~~~~~~~~~~~~~~~~~
\nonumber\\
\e^{\a_1..\a_{16}} (\g^{a_1a_2a_3}_{\a_1\a_2}\g^{b_1b_2b_3}_{\a_3
\a_4} )(\g^{c_1c_2c_3}_{\a_5\a_6}\g^{d_1d_2d_3}_{\a_7 \a_8} )\dots
(\g^{g_1g_2g_3}_{\a_{13}\a_{14}}\g^{h_1h_2h_3}_{\a_{15} \a_{16}}
)\times
\nonumber\\
\car_{a_1a_2a_3b_1b_2b_3}\car_{c_1c_2c_3d_1d_2d_3}\dots
\car_{g_1g_2g_3h_1h_2h_3}\eea where we have defined {\footnote {In a
previous draft, we had used the Riemann tensor instead of the Weyl
tensor. We would like to thank the authors of \cite{deHaro:2002vk}
for pointing out this error.}} \be
 \car_{abcdef} =\frac{1}{16}(g_{ad} c_{bcef} -
\frac{i}{6} D_b g_{acdef}) -\frac{1}{1536} (3 g_{bafmn}
g_{ced}{}^{mn} - g_{abcmn} g_{def}{}^{mn}) \ee

The fermionic terms can be written down similarly.

\vskip 1 cm

This procedure can be generalized by considering any action of the
form (\ref{action}), with W being replaced by any function of V, and
with the first two components of $\Delta$ given in equation
(\ref{Deltaeqn}). Then our analysis proceeds unchanged, and we can
shown that all variations with one fermion field and
$\del{\tau}=\l=a_2=0$ cancel. This allows us to generate a large set
of actions involving $R,g_5$ and $\psi$ which have at least a partial
cancellation of the supersymmetry variations.

\section{Discussion and a conjecture}

Let us now summarize our results.

We have found the quartic action of type IIB including the moduli
dependence in equation (\ref{linaction}). Supersymmetry variations of
this action will produce a large number of variations; the terms
which are quartic in the fields will cancel.

We then extended the analysis beyond the quartic level. We restricted
ourselves to terms involving the curvature $R$ and five-form field
strength $g_5$. To cancel the variations coming from these terms we
also considered terms containing the curvature, the five form field
strength and in addition, two gravitino fields.

These terms were found using the superfield construction of
\cite{HW,deHaro:2002vk}. In terms of this superfield, the terms
discussed in the paragraph above can all be generated by the action
(\ref{action}) where the components of the measure are given in
equation (\ref{Deltaeqn}). In particular, all bosonic terms involving
$R$ and $g_5$ are given by the action (\ref{bosact}). This confirms
the results of \cite{previous}.

\vskip 1 cm

In hindsight, it is clear that the superfield construction cannot
give us the full action. The action (\ref{act1})  depends on an
arbitrary function W, while the real action is completely determined
by supersymmetry up to an overall factor. In other words, we normally
use superfields in order to be able to construct the most general
supersymmetric action. But the action up to eight derivatives is
unique, so the superfield construction (\ref{act1}) will not work.
(It may be possible to use a non-scalar superfield, in which case the
action would not be of the form (\ref{act1}). We would like to thank
the authors of \cite{deHaro:2002vk} for bringing this possibility to
our attention.)

Here we argued that despite this failure, the superfield indeed
reproduces a subset of terms; in particular all terms with only $R$
and $g_5$ can be reproduced. It is now natural to ask if we can
extend this approach to obtain the remaining terms in the action.

\vskip 1 cm

First, we note that the vanishing of the obstruction
(\ref{obstruction}) does not require us to impose $a_2=0$. This
suggests the

\vskip 1 cm

{\it Conjecture: A superfield $\Delta$ can be constructed such the
subset of the supersymmetry variations of the action (\ref{act1})
which have one fermion field and which satisfy
$\del{\tau^*}=\del{\tau}=\l=0$ all vanish.}

\vskip 1 cm

Proving this conjecture would require us to continue generating
further projections of $\D$ by the procedure outlined in
\cite{deHaro:2002vk}, and reviewed above.  We will leave the proof or
disproof of this conjecture to future work.

It does not seem possible to extend the scope of the superfield
further than this conjectured extension. The apparently insuperable
obstruction is the fact the moduli-dependent coefficients are
non-holomorphic. It is hard to see how any integral of holomorphic
field can produce such coefficients. The integral over superspace
will generically produce moduli dependent coefficients $W[v,u^*]$,
which cannot correspond to the required coefficients
$f^{(w,-w)}(\tau,\tau^*)$.

This suggests that we will have to give up the idea of obtaining the
coefficients from a superfield construction  and instead consider an
action of the form \bea S_8=\int d^{10}x f^{(w,-w)}(\tau,\tau^*)\int
d^{16}\theta \Delta W[V,U^*] \eea The cancellation of terms coming
from the variations of the moduli-dependent coefficients will not
occur, and accordingly, the supersymmetry variations of this action
will have uncancelled terms containing either a derivative acting on
a scalar ($\del{\tau^*}$ or $\del{\tau}$) or a dilatino factor. Hence
it would seem impossible that the entire action can be constructed
using superfields.

\vskip 1 cm

It is possible that to find the full action, one will have to modify
the superfield approach drastically. In particular, when we
constructed the superfield, we imposed certain constraints on the
torsion and curvature. These constraints are required in order to
reduce the large number of fields in the superfield. Now it is very
possible that the $\alpha'$ corrections modify these constraints as
well. The corrected constraints will produce a superfield different
from the one we considered, and may lead to a supersymmetric action
\cite{Cederwall:2000ye}. Unfortunately, it is a very difficult task
to find the correct constraints with no additional help. The
construction given in this paper may be of help in finding these
constraints.

\vskip 1 cm

{\bf Acknowledgements}: We are grateful to N.~Berkovits, S.~Deser,
S.~de Haro, M.~Green, P.~Howe, A.~Sinkovics and K.~Skenderis for
useful comments.

The author is supported in part by NSF Grant PHY-0354993.

\end{document}